\documentclass[10pt,hidelinks,conference]{IEEEtran}
\IEEEoverridecommandlockouts

\usepackage{cite}
\usepackage{amsmath,amssymb,amsfonts}
\usepackage{graphicx}
\usepackage{textcomp}
\usepackage[dvipsnames]{xcolor}
\usepackage{orcidlink}
\usepackage[acronym,shortcuts]{glossaries}
\usepackage{bm}
\usepackage{caption}
\usepackage{subcaption}
\usepackage{relsize}
\usepackage{soul}
\usepackage{algorithm, algpseudocode}
\usepackage{outlines}
\usepackage{lipsum}
\usepackage{scalefnt}
\usepackage{dblfloatfix}
\usepackage[bottom]{footmisc}
\newcommand{\trans}[0]{^{\mathsf{T}}}
\newcommand{\transs}[0]{^{\!\mathsf{T}}}

\newacronym{WSN}{WSN}{wireless sensor network}
\newacronym{WLS}{WLS}{weighted least squares}
\newacronym{DAC}{DAC}{divide and conquer}
\newacronym{TDOA}{TDOA}{time difference of arrival}
\newacronym{TOA}{TOA}{time of arrival}
\newacronym{IMU}{IMU}{inertial measurement unit}
\newacronym{MFB}{MFB}{matched filter bound}
\newacronym{BP}{BP}{belief propagation}
\newacronym{RMSE}{RMSE}{root-mean-squared-error}
\newacronym{VR}{VR}{virtual reality}
\newacronym{XR}{XR}{extended reality}
\newacronym{SotA}{SotA}{state-of-the-art}
\newacronym{MSE}{MSE}{mean-squared-error}
\newacronym{PDF}{PDF}{probability density function}
\newacronym{SGA}{SGA}{scalar Gaussian approximation}
\newacronym{ML}{ML}{machine learning}
\newacronym{IC}{IC}{interference cancellation}
\newacronym{GaBP}{GaBP}{Gaussian belief propagation}
\newacronym{RBL}{RBL}{rigid body localization}
\newacronym{AWGN}{AWGN}{additive white Gaussian noise}
\newacronym{3D}{3D}{three-dimensional}
\newacronym{IoT}{IoT}{Internet-of-Things}
\newacronym{i.i.d.}{i.i.d.}{independent and identically distributed}
\newacronym{SVD}{SVD}{singular value decomposition}

\title{Belief Propagation-based Rotation and Translation \\[-0.2ex] Estimation for Rigid Body Localization}

\author{\IEEEauthorblockN{Volodymyr~Vizitiv\textsuperscript{\orcidlink{0009-0002-9451-8216}}, Hyeon Seok Rou\textsuperscript{\orcidlink{0000-0003-3483-7629}}, Niclas~F\"uhrling\textsuperscript{\orcidlink{0000-0003-1942-8691}}, and Giuseppe Thadeu Freitas de Abreu\textsuperscript{\orcidlink{0000-0002-5018-8174}}}
\IEEEauthorblockA{\textit{School of Computer Science and Engineering, Constructor University, 28759 Bremen, Germany} \\
$[$vvizitiv, hrou, nfuehrling, gabreu$]$@constructor.university}\\[-5ex]}

\begin{document}
\maketitle

\begin{abstract}
\begin{outline}
We propose a novel solution to the \ac{RBL} problem, in which the \ac{3D} rotation and translation is estimated by only utilizing the range measurements between the wireless sensors on the rigid body and the anchor sensors.
Given the prior knowledge of the absolute sensor positions, by leveraging a linearized \ac{RBL} transformation model with small-angle approximations, the proposed bivariate \ac{GaBP} is designed to directly estimate the \ac{3D} rotation angles and translation distances, with an \ac{IC} refinement step to further improve the angle estimation performance.
The effectiveness of the proposed method is verified via numerical simulations, highlighting the superior performance of the proposed method against the \ac{SotA} techniques for the rotation and translation estimation performance.

\end{outline}
\end{abstract}
\glsresetall

\begin{IEEEkeywords}
\Ac{RBL} and tracking, range-based positioning, \ac{GaBP}.
\\[-4ex]
\end{IEEEkeywords}

\glsresetall

\section{Introduction}
\label{sec:introduction}
\vspace{-1.5ex}

Recent years have seen noteworthy advancements in wireless sensor technology capable of both wireless communications and environment parameter detection (\textit{i.e.,} temperature, illuminance, signals), directing significant attention towards \acp{WSN} with the inherent implications for applications requiring monitoring and control, such as in smart factories and \ac{IoT} \cite{Jamshed_SJ22, Kandris_ASI20}.
%

In many of the key \ac{WSN} applications, including logistics, healthcare, and security, the accurate geographic location information of the sensors is essential, facing a unique challenge in that the sensor position is not a locally measurable instantaneous parameter.
Consequently, the problem is already well-identified and extensively studied as the sensor localization problem \cite{GustafssonSPM2005}.
However, next-generation technologies and applications such as \ac{VR}, \ac{XR}, robotics, and autonomous vehicles, require not only the precise location information of the sensors, but also the orientation of the sensor network associated to a given body or object \cite{Yang_JMD09,Lee_Sensor18,Whittaker_IEEE06}.
This challenge has gained increasing popularity and is referred to as the \ac{RBL} problem \cite{Chepuri_TSP14,Yu_ITJ23}, where the sensor network is defined as a \textit{rigid} conformation, whose translation and rotation (\textit{i.e.,} orientation) must be estimated for single or multiple \textit{bodies}.

Several effective strategies have been developed to address the \ac{RBL} problem, for example, computer vision-based techniques for feature extraction and posture estimation \cite{Chaoyi_ICASSP21, Xiang_arxiv17} often requiring high volumes of image/video data and high-complexity methods involving \ac{ML}, or \ac{IMU}-based orientation and position estimation techniques leveraging the internal information of the accelerometers, gyroscopes, and magnetometers \cite{AghiliTM2013,Zhao_JPCS18} often requiring costly hardware, precise sensor calibration, and aid of external radio technology.

An alternative approach with relatively small processing requirement and simple hardware, aims to exploit the range measurements between the sensors on the rigid body and the set of \textit{anchor} sensors of known positions 
\cite{AlcocerCDC2008,SandPWCS2014}, \textit{i.e.,} sensors on static infrastructure of the environment, which can be obtained from the \ac{TOA} or \ac{TDOA} data often already available from the communications functionality of the wireless sensors.
In such methods, the initial estimation of the sensor position is performed with the range measurement data without the sensor conformation, followed by the extraction of the translation and rotation parameters (translation distance and rotation angles associated with each axis) subject to the rigid body constraints \cite{BeckTSP2008,ChanTSP1994}.
To elaborate on a \ac{SotA} example, the method in \cite{ChenTSP2015} leverages the \ac{DAC} approach \cite{AbelTAES1990} to estimate the sensor positions from the range measurements, then extracts the rotation and translation parameters via \ac{SVD}-based analysis.
Finally, a conformation-based refinement based on the Euler angles formulation and \ac{WLS} is performed on the obtained rotation and translation estimates.

In light of the above, we propose a novel \ac{RBL} algorithm which is capable of estimating the translation distance and rotation angle from the sensor range measurements and the rigid body conformation, enabled by a linearized reformulation of the system model and a tailored design of two low-complexity \ac{GaBP} \cite{bickson2009gaussian} estimators. 
%
%
The proposed method is shown to outperform the \ac{SotA} two-stage \ac{RBL} methods in the translation and rotation estimation performance, while retaining the low computational complexity that is shown to be linear on the number of sensors.

The remainder of the article is structured as follows: Section \ref{sec:system_model} describes the system model and formulates the fundamental \ac{RBL} estimation problem, Section \ref{sec:reformulation} presents the proposed linearized reformulation leveraging small angle approximation and the \ac{RBL} conformation constraints, Section \ref{sec:proposed} elaborates the derivation of the message passing rules for the proposed \ac{GaBP} estimator for the translation and rotation parameters, and finally Section \ref{sec:performance_analysis} compares the performance of the proposed method against the \ac{SotA} two-stage method \cite{ChenTSP2015} via numerical simulations.

\newpage
\vspace{-1ex}
\section{Rigid Body Localization System Model}
\label{sec:system_model}
\vspace{-1ex}
\subsection{Rigid Body System Model}
\label{sec:RBL_model}

\begin{figure}[b]
\centering
\vspace{-3ex}
\includegraphics[width=0.85\columnwidth]{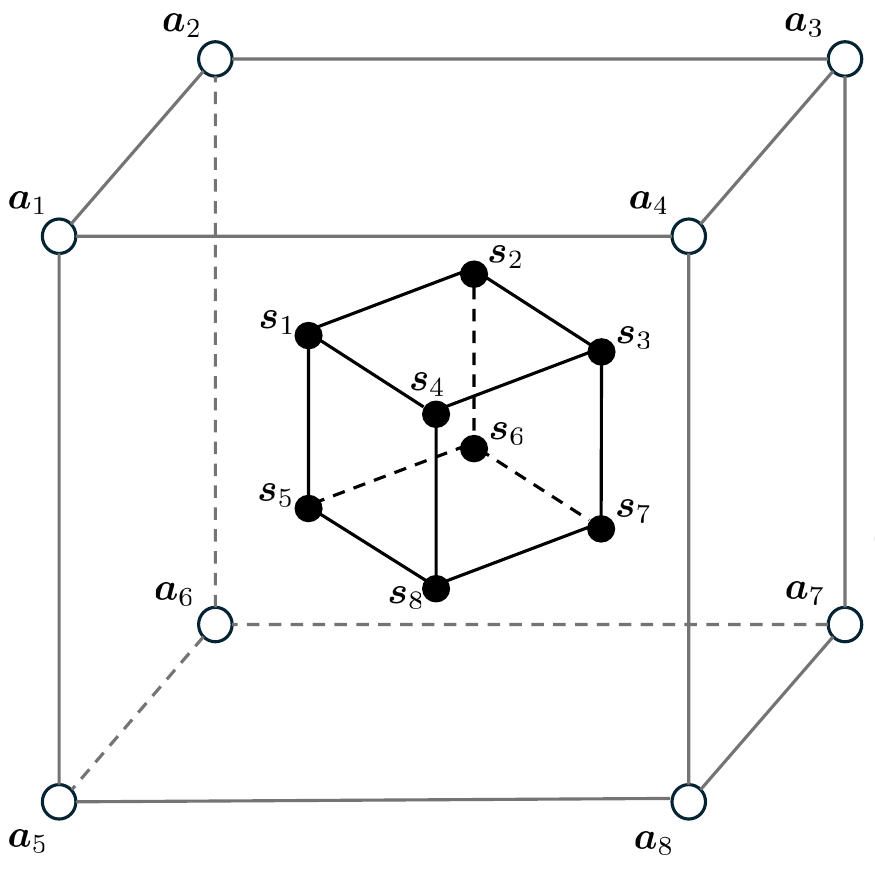}    
\vspace{-1ex}
\caption{An illustration of the rigid body sensor and anchor structure, exemplified by a cubic rigid body ($N = 8$) surrounded by a cubic deployment of anchors ($M = 8$).}
\label{fig:sys_mod_plot}
\vspace{-3ex}
\end{figure}

Consider a scenario where a rigid body consisting of $N$ sensors is surrounded by a total of $M$ anchor sensors (hereafter referred simply as anchors), as illustrated in Figure \ref{fig:sys_mod_plot}.
The sensors are described by a $3 \times 1$ vector consisting of the \textit{x-,y-,z}-coordinates in the \ac{3D} Euclidean space, denoted by $\boldsymbol{c}_n \in \mathbb{R}^{3\times 1}$ for $n=\{1, \ldots, N\}$ and $\boldsymbol{a}_m \in \mathbb{R}^{3\times 1}$ for $m=\{1, \ldots, M\}$, respectively for the rigid body sensors and anchors.
The initial sensor structure in the rigid body is consequently defined by the conformation matrix $\boldsymbol{C}=\left[\boldsymbol{c}_{1}, \boldsymbol{c}_{2}, \ldots, \boldsymbol{c}_{N}\right] \in \mathbb{R}^{3 \times N}$ at the reference frame (local axis) of the rigid body.

A transformation of the rigid body in \ac{3D} space can be fully defined by a translation and rotation, respectively described by the translation vector $\boldsymbol{t} \triangleq [t_x, t_y, t_z]\trans \in\mathbb{R}^{3\times 1}$ consisting of the translation distances in each axis, and a \ac{3D} rotation matrix\footnote{The rotation matrix $\boldsymbol{Q}$ is part of the special orthogonal group such that $SO(3)=\{\boldsymbol{Q} \in \mathbb{R}^{3 \times 3}: \boldsymbol{Q}\trans \boldsymbol{Q} = \mathbf{I}, ~\mathrm{det}(\boldsymbol{Q})=1\}$ \cite{Diebel2006RigidBodyAttitude}.} $\boldsymbol{Q} \in \mathbb{R}^{3\times 3}$ given by \vspace{-1ex}
\begin{align}
\label{eq:rotation_matrix}
\bm{Q} \triangleq\! \text{\scalebox{0.875}{$\overbrace{\left[
\begin{array}{@{}c@{\;\,}c@{\;\,}c@{}}
\cos\theta_z&-\sin\theta_z& 0\\
\sin\theta_z& \cos\theta_z& 0\\
0 	    & 0           & 1\\
\end{array}\right]}^{\triangleq \,\bm{Q}_z \in\, \mathbb{R}^{3\times 3}}\!\cdot\!
\overbrace{\left[
\begin{array}{@{}c@{\;\,}c@{\;\,}c@{}}
\cos\theta_y & 0           & \sin\theta_y\\
0			& 1			  & 0\\
-\sin\theta_y& 0 		  & \cos\theta_y\\
\end{array}\right]}^{\triangleq \,\bm{Q}_y \in\, \mathbb{R}^{3\times 3}}\!\cdot\!
\overbrace{\left[
\begin{array}{@{\,}c@{\;\,}c@{\;\,}c@{\!}}
1 			& 0			  & 0\\
0			& \cos\theta_x& -\sin\theta_x\\
0			& \sin\theta_x& \cos\theta_x\\
\end{array}\right]}^{\triangleq \bm{Q}_x \in\, \mathbb{R}^{3\times 3}}$}}\!, \\[-5ex] \nonumber 
\end{align}
where $\bm{Q}_{x}, \,\bm{Q}_{y}, \,\bm{Q}_{z} \in \mathbb{R}^{3\times 3}$ are the roll, pitch, and yaw rotation matrices about the \textit{x-,y-,z-}axes by rotation angles of $\theta_x, \theta_y, \theta_z \in [-180^\circ, 180^\circ]$ degrees, respectively.

\begin{figure}[t]
\centering
\includegraphics[width=0.95\columnwidth]{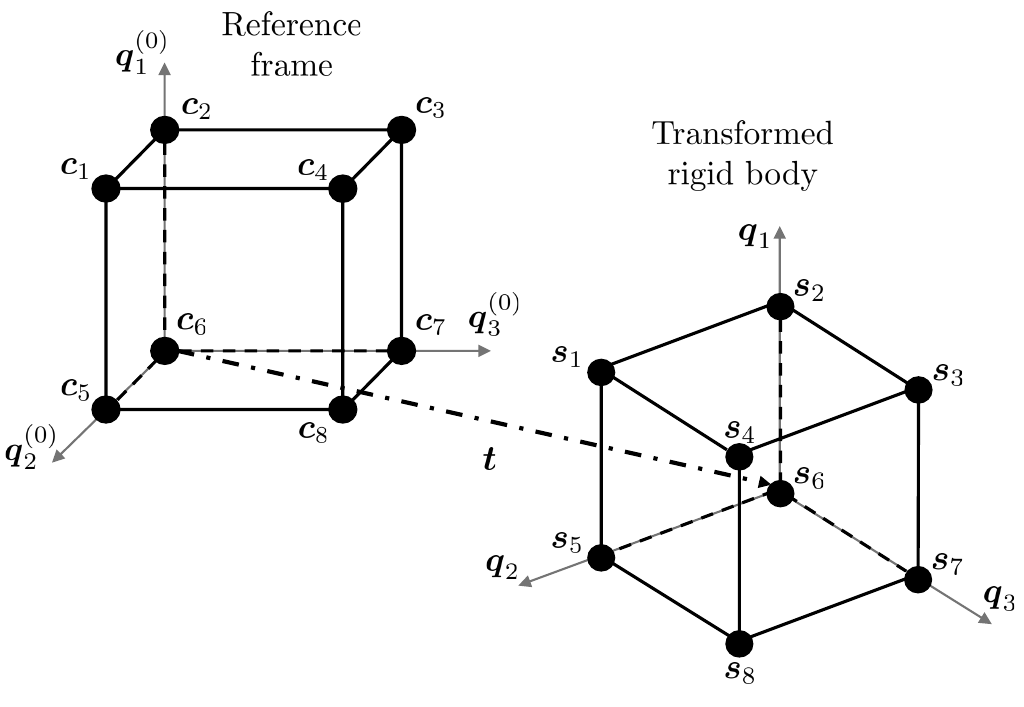}  
\vspace{-2ex}
\caption{An illustration of the rigid body transform composed of a \ac{3D} rotation $\boldsymbol{Q}$ and translation $\boldsymbol{t}$, from the reference frame\protect\footnotemark ~to the transformed sensor positions determined by equation \eqref{eq:basic_model_RB}.}
\label{fig:RB_trans_plot}
\vspace{-3.5ex}
\end{figure}

In light of the above, the transformed coordinates of the $n$-th sensor after the rotation and translation is described by
\begin{equation}
\label{eq:basic_model_RB}
\boldsymbol{s}_{n} =\boldsymbol{Q} \boldsymbol{c}_{n}+\boldsymbol{t} \in \mathbb{R}^{3 \times 1},
\end{equation}
which is applied identically to all $N$ sensors of the rigid body, as illustrated in Figure \ref{fig:RB_trans_plot}.

\vspace{-2ex}

\subsection{Position-based System Model}

\footnotetext{The reference frame $\boldsymbol{Q}^{(0)}$ may already be at an non-origin position, but since a rigid body rotation is only relative between the initial frame and the new orientation, the reference frame can be considered to be at the origin $\boldsymbol{Q}^{(0)} \triangleq \mathbf{I}_{3\times 3}$ without loss of generality.}

In this article, \ac{RBL} is performed using the pairwise range measurement information between the anchors and sensors, which is assumed to be available and described by
\begin{equation}
\label{eq:range_model}
\tilde{d}_{m,n} = d_{m,n} + w_{m,n} = \left\|\boldsymbol{a}_{m} - \boldsymbol{s}_{n}\right\|_{2} + w_{m,n} \in \mathbb{R},
\end{equation}
where $d_{m,n} \triangleq \left\|\boldsymbol{a}_{m} - \boldsymbol{s}_{n}\right\|_{2} \in \mathbb{R}$ is the true Euclidean distance between $m$-th anchor and $n$-th sensor, and $w_{m,n} \sim \mathcal{N}(0,  \sigma_w^2)$ is the \ac{i.i.d.} \ac{AWGN} of the range measurement with noise variance $\sigma_w^2$.

Following the above, the squared information of the range measurement is given by
\begin{equation}
\label{eq:sq_range_meas}
\tilde{d}_{m,n}^2 = \left\|\boldsymbol{a}_{m} - \boldsymbol{s}_{n}\right\|_{2}^2 + 2d_{m,n}w_{m,n} + w_{m,n}^2  \in \mathbb{R},
\end{equation}
which can be reformulated as the \textit{composite noise} $\xi_{n} \in \mathbb{R}$ as
\begin{equation}
\label{eq:pos_lin_eq}
\xi_{m,n} = \tilde{d}_{m,n}^{2} - \left\|\boldsymbol{a}_{m}\right\|^{2}_2 -\left\|\boldsymbol{s}_{n}\right\|^{2}_2 +2 \boldsymbol{a}_{m}\trans \boldsymbol{s}_{n} \approx 2d_{m,n}w_{m,n},
\end{equation}
where the second-order noise term $w_{m,n}^2$ is considered negligible and neglected \cite{ChenTSP2015,HoTSP2004}.

Stacking equation \eqref{eq:pos_lin_eq} for all $M$ anchors and reformulating as a linear system on the $n$-th unknown sensor variable yields \vspace{-3ex}
\begin{align}
\boldsymbol{y}_{n} \!\!&\triangleq\!\!\!
\begin{bmatrix}
\!\tilde{d}_{1,n}^{2} \!\!-\! \left\|\boldsymbol{a}_{1}\right\|^{2}_2 \!\\
\!\vdots \\
\tilde{d}_{M,n}^{2} \!\!-\! \left\|\boldsymbol{a}_{M}\right\|^{2}_2 \\
\end{bmatrix} 
\!\!\!=\!\!\!
\underbrace{
\begin{bmatrix}
\!-2 \boldsymbol{a}_{1}\trans, &\!\!\!\! 1\; \\
\!\vdots &\!\!\!\! \vdots~\\
\!-2 \boldsymbol{a}_{M}\trans, &\!\!\!\! 1\; \\
\end{bmatrix}}_{\triangleq \,\boldsymbol{G} \,\in\, \mathbb{R}^{M \!\times\!4}}
\!\!\!\!\!\!\!\overbrace{\!
\begin{bmatrix}
\!\boldsymbol{s}_n \!\\[1ex]
\!||\boldsymbol{s}_n||_2^2\!
\end{bmatrix}\!}^{ ~~~\triangleq \, \boldsymbol{x}_n \, \in \, \mathbb{R}^{4 \!\times\!1}}
\!\!\!\!\!\!+\!\!\!\!\!\!
\underbrace{\!
\begin{bmatrix}
\!\xi_{1,n}\!\\
\!\vdots\!\\
\!\xi_{M,n}\!\\
\end{bmatrix}
\!}_{\triangleq \,\boldsymbol{\xi}_n \,\in\, \mathbb{R}^{M \!\times\!1}}\!\!\!\!\!\!\! \in \!\mathbb{R}^{M \times 1}\!, \nonumber \\[-4ex] 
\label{eq:linear_sys}
\end{align}
where $\boldsymbol{y}_{n} \in \mathbb{R}^{M \times 1}$ and $\boldsymbol{G}$ $\in \mathbb{R}^{M \times 4}$ are respectively the observed data vector and effective channel matrix constructed from the measured ranges and anchor positions, $\boldsymbol{x}_n \in \mathbb{R}^{4 \times 1}$ is the unknown sensor position vector, and $\boldsymbol{\xi}_{n} \in \mathbb{R}^{M \times 1}$ is the vector of composite noise variables from equation \eqref{eq:pos_lin_eq}.

The linear system in equation \eqref{eq:linear_sys} can be leveraged for the estimation of the unknown sensor coordinate vector $\boldsymbol{s}_n$ and sensor position norm $||\boldsymbol{s}||_2^2$ in $\boldsymbol{x}_n$, from which equation \eqref{eq:basic_model_RB} can be invoked for the translation and rotation extraction via Procrustes analysis or other classical algorithms \cite{EggertMVA1997}.
\vspace{-1ex}
\subsection{Transformation Parameter-based System Model}
\label{sec:reformulation}

In this section, the fundamental system of equation \eqref{eq:linear_sys} is reformulated to express the system directly in terms of the \ac{RBL} transformation parameters, \textit{i.e.,} the \ac{3D} rotation angles $\boldsymbol{\theta} \triangleq [\theta_x, \theta_y, \theta_z]\trans \in \mathbb{R}^{3 \times 1}$ and translation vector $\boldsymbol{t}$ \cite{ZhaMRBL2021}.
First, small-angle approximation\footnote{For practical rigid body tracking applications, subsequent transformation estimations can be assumed to be performed within a sufficiently short time period, such that the change in rotation angle is small.} \cite{Diebel2006RigidBodyAttitude} of the rotation matrix in equation \eqref{eq:rotation_matrix} is obtained with $\cos\theta \approx 1$ and $\sin\theta \approx \theta$, as
\begin{eqnarray}
\label{eq:q_small_angle}
~~~~\bm{Q} \approx\!\!\left[\begin{array}{ccc}
1 \!&\! \theta_z \!&\! -\theta_y \\[-0.5ex]
-\theta_z \!&\! 1 \!&\! \theta_x \\[-0.5ex]
\theta_y \!&\! -\theta_x \!&\! 1
\end{array}\right] \in \mathbb{R}^{3 \times 3},
\end{eqnarray}
which in turn can be vectorized into a linear system directly in terms of the Euler angles \cite{ChenTSP2015} as \vspace{-0.5ex}
\begin{align}
\label{eq:q_vec}
\mathrm{vec}(\boldsymbol{Q}) &= \boldsymbol{\gamma} + \boldsymbol{L} \boldsymbol{\theta} \\[-1ex]
&= \overbrace{
\begin{bmatrix}
1 & 0 & 0 & 0 & 1 & 0 & 0 & 0 & 1
\end{bmatrix}\transs}^{\triangleq\, \boldsymbol{\gamma} \, \in \, \mathbb{R}^{9 \times 1}} \nonumber \\
& + \underbrace{
\begin{bmatrix}
0 &  1 \!\!&\! 0 \!\!\!&\!\! -1 \!\!& 0 & 0 & 0 \!\!\;\!\!& 0 \!\!\;\!\!& 0 \\ 
0 &  0 \!\!&\! -1 \!\!\!&\!\! 0 \!\!& 0 & 0 & 1 \!\!\;\!\!& 0 \!\!\;\!\!& 0 \\ 
0 &  0 \!\!&\! 0 \!\!\!&\!\! 0 \!\! & 0 & 1 & 0 \!\!\;\!\!& -1 \!\!\;\!\!& 0 \\ 
\end{bmatrix}\transs\!\!\!}_{\triangleq\, \boldsymbol{L} \, \in \, \mathbb{R}^{9 \times 3}} 
\!\cdot\!
\begin{bmatrix}
\theta_x \\ \theta_y \\ \theta_z
\end{bmatrix}\!\in \mathbb{R}^{9 \times 1}. \nonumber
\end{align}

Then substituting equation \eqref{eq:q_vec} into equations \eqref{eq:basic_model_RB} and \eqref{eq:pos_lin_eq} and rearranging the terms yields the alternate representation of the composite noise as
\begin{equation}
\begin{aligned}
~~~~~~\xi_{n} &= \tilde{d}_{m,n}^{2}-\left\|\boldsymbol{a}_{m}\right\|^{2}_2 - \left\|\boldsymbol{s}_{n}\right\|^{2}_2 + 2\!\left[\boldsymbol{c}_{n}\trans \otimes \boldsymbol{a}_{m}\trans \right]\!\boldsymbol{\gamma}\\
&\quad+2\!\left[\boldsymbol{c}_{n}\trans \otimes \boldsymbol{a}_{m}\trans\right]\!\boldsymbol{L}\boldsymbol{\theta} +2 \boldsymbol{a}_{m}\trans \boldsymbol{t} ~~~~~~~~~~~~~~~~ \in \mathbb{R},\!\! \label{eq:delta_lin_eq}
\end{aligned}
\end{equation}
where the matrix product vectorization identity $\mathrm{vec}(\mathbf{X Y Z}) = (\mathbf{Z}\trans \!\otimes\! \mathbf{X}) \mathrm{vec}(\mathbf{Y})$ has been used, with Kronecker product $\otimes$.

In light of the above, the fundamental system can be rewritten leveraging the linearization of equation \eqref{eq:delta_lin_eq},
\begin{subequations}
\label{eq:delta_t_lin_syst}
\begin{equation}
\boldsymbol{z}_{n} = \boldsymbol{H}^{\theta} \!\cdot\! \boldsymbol{\theta} + \boldsymbol{H}^{t} \!\cdot\! \boldsymbol{t} + \boldsymbol{\xi}_{n} \in \mathbb{R}^{M \times 1}, \vspace{-1ex}
\label{eq:delta_t_lin_syst_a}
\end{equation}
with
\begin{align}
\boldsymbol{z}_{n} \!=\!\! \left[\!\!\begin{array}{c}
\tilde{d}_{1,n}^{2}-\left\|\boldsymbol{a}_{1}\right\|^{2}_2- \left\|\boldsymbol{s}_{n}\right\|^{2}_2 + 2\left[\boldsymbol{c}_{n}\trans \otimes \boldsymbol{a}_{1}\trans \right]\!\boldsymbol{\gamma} \\
\vdots \\
\tilde{d}_{M,n}^{2}-\left\|\boldsymbol{a}_{M}\right\|^{2}_2 - \left\|\boldsymbol{s}_{n}\right\|^{2}_2 + 2\left[\boldsymbol{c}_{n}\trans \otimes \boldsymbol{a}_{M}\trans \right]\!\boldsymbol{\gamma}
\end{array}\!\!\right] \!\! \in \mathbb{R}^{M \times 1}, \nonumber \\[-3.25ex] \\[-4.5ex] \nonumber
\end{align}
\begin{align}
\boldsymbol{H}^{\theta} \!=\!\! \left[\begin{array}{c}
\!\!\!\!\!-2\!\left[\boldsymbol{c}_{i}\trans \otimes \boldsymbol{a}_{1}\trans \right]\!\boldsymbol{L}\!\!\!\\
\vdots   \\[0.5ex]
\!\!\!\!-2\!\left[\boldsymbol{c}_{i}\trans \otimes \boldsymbol{a}_{M}\trans \right]\!\boldsymbol{L}\!\!\!
\end{array}\right] \!\! \in \!\mathbb{R}^{M \times 3}\!, 
~ \boldsymbol{H}^{t} \!=\!\! \left[\begin{array}{c}
\!\!\!\!-2 \boldsymbol{a}_{1}\trans\!\!\!\!  \\
\vdots \\
\!\!\!\!-2 \boldsymbol{a}_{M}\trans\!\!\! 
\end{array}\right]\!\! \in \!\mathbb{R}^{M \times 3}\!, \nonumber \\[-3.25ex] \\[-3ex] \nonumber
\end{align}
\end{subequations}
where $\boldsymbol{z}_{n} \in \mathbb{R}^{M \times 1}$ is the effective observed data vector, and $\boldsymbol{H}^{\theta} \in \mathbb{R}^{M \times 3}$ and $\boldsymbol{H}^{t} \in \mathbb{R}^{M \times 3}$ are respectively the effective channel matrices for the unknown rotation and translation variables, and $\boldsymbol{\xi}_{n} \in \mathbb{R}^{M \times 1}$ is the vector of composite noise variables from equation \eqref{eq:pos_lin_eq}.

%

\section{Proposed \ac{RBL} Method}
\label{sec:proposed}

In light of the system model derived in Section \ref{sec:system_model}, a low-complexity  transformation parameter estimator for \ac{RBL} is proposed by a tailored message passing algorithm under the \ac{GaBP}  framework.
Given the absolute sensor coordinate information obtained by solving the position-explicit system in equation \eqref{eq:linear_sys} using existing methods such as closed-form two-stage method \cite{MaICASSP2011} or any other \ac{SotA} range-based technique, a proper \ac{GaBP} is derived on the rigid body transformation parameter-explicit system in equation \eqref{eq:delta_t_lin_syst}, to obtain the final estimate of the rotation angle $\boldsymbol{\theta}$ and translation $\boldsymbol{t}$.


\subsection{Bivariate GaBP for Transformation Parameter Estimation}
\label{sec:del_t_est_sol}

\textit{\textbf{Remark:} As the \ac{GaBP} derivation for each $n$-th linear system corresponding to the $n$-th sensor node $\forall\,n \in \{1,\ldots,N\}$ are identical, the subscript $n$ is omitted from onwards.}

As can be seen in the reformulated system of the transformation parameter estimation in equation \eqref{eq:delta_t_lin_syst_a}, there exist two sets of variables $\theta_k$ with $k \in \{1,\ldots,K\}$ and $t_\ell$ with $\ell \in \{1,\ldots,K\}$, whose soft-replicas of the $k$-th, $\ell$-th elements via the $m$-th observation is denoted by $\hat{\theta}_{m,k}^{[\lambda]}$, $\hat{t}_{m,\ell}^{[\lambda]}$, respectively.

Firstly, the soft-\ac{IC} is performed on the observed information respectively for the angle and translation variables as
\begin{subequations}
\begin{equation}\label{eq:del_IC}
\resizebox{0.48 \textwidth}{!}{$
\begin{aligned}
\tilde{z}_{m,k}^{\theta[\lambda]} &= z_{m} - \sum_{i \neq k} h^{\theta}_{m,i}\hat{\theta}_{m,i}^{[\lambda]} - \sum_{i = 1}^{K} h^t_{m,i}\hat{t}_{m,i}^{[\lambda]}, \\
&= h^{\theta}_{m,k}\theta_{k} + \sum_{i \neq k} h^{\theta}_{m,i}(\theta_{i} - \hat{\theta}_{m,i}^{[\lambda]}) + \sum_{i = 1}^{K} h^{t}_{m,i}(t_{i} - \hat{t}_{m,i}^{[\lambda]})+ \xi_m, \\[-11ex]
\end{aligned}
$}
\end{equation}

\vspace{4ex}

\begin{equation}\label{eq:t_IC}
\resizebox{0.48 \textwidth}{!}{$
\begin{aligned}
\tilde{z}_{m,\ell}^{t[\lambda]} &= z_{m} - \sum_{i = 1}^{K} h^{\theta}_{m,i}\hat{\theta}_{m,i}^{[\lambda]} - \sum_{i \neq \ell}  h^{t}_{m,i}\hat{t}_{m,i}^{[\lambda]}, \\
&=  h^{t}_{m,\ell}t_{\ell} + \sum_{i = 1}^{K}  h^{\theta}_{m,i}(\theta_{i} - \hat{\theta}_{m,i}^{[\lambda]}) + \sum_{i \neq \ell}  h^{t}_{m,i}(t_{i} - \hat{t}_{m,i}^{[\lambda]}) + \xi_m. \\[-12ex]
\end{aligned}
$}
\end{equation}
\vspace{8ex}
\label{eq:del_t_soft_IC}
\end{subequations}

In turn, the conditional \acp{PDF} of the soft-\ac{IC} symbols are given by
\begin{equation}
\label{eq:del_t_cond_PDF}
\begin{aligned}
p_{\tilde{\mathrm{z}}_{m,k}^{\theta[\lambda]} \mid \mathrm{\theta}_{k}}(\tilde{z}_{m,k}^{\theta[\lambda]}|\theta_{k}) &\propto \mathrm{exp}\left[ -\frac{|\tilde{z}_{m,k}^{\theta[\lambda]} - h^{\theta}_{m,k} \theta_{k}|^2}{\big(\sigma_{m,k}^{\theta[\lambda]}\big)^2} \right], \\[1ex]
p_{\tilde{\mathrm{z}}_{m,\ell}^{t[\lambda]} \mid \mathrm{t}_{\ell}}(\tilde{z}_{m,\ell}^{t[\lambda]}|t_{\ell}) &\propto \mathrm{exp}\left[ -\frac{|\tilde{z}_{m,\ell}^{t[\lambda]} - h^{t}_{m,\ell} t_{\ell}|^2}{\big(\sigma_{m,\ell}^{t[\lambda]}\big)^2} \right],
\end{aligned}
\end{equation}
with conditional variances \vspace{-1ex}
\begin{subequations}
\begin{align}
\nonumber
\big(\sigma_{m,k}^{\theta[\lambda]}\big)^2 &= \sum_{i \neq k} \left|h^{\theta}_{m,i}\right|^2\psi_{m,i}^{\theta[\lambda]} + \sum_{i = 1}^{K} \left|h^{t}_{m,i}\right|^2\psi_{m,i}^{t[\lambda]} + N_{0} \in \mathbb{R},\\[-3ex]\label{eq:del_theta_var} \\
\big(\sigma_{m,\ell}^{t[\lambda]}\big)^2 &= \sum_{i = 1}^{K} \left|h^{\theta}_{m,i}\right|^2\psi_{m,i}^{\theta[\lambda]} + \sum_{i \neq \ell} \left|h^{t}_{m,i}\right|^2\psi_{m,i}^{t[\lambda]} + N_{0} \in \mathbb{R}, \nonumber\\[-3ex]\label{eq:del_t_var}
\end{align}
\label{eq:del_t_theta_var}
\end{subequations}
\vspace{-1ex}

\noindent where $\psi_{m,k}^{\theta[\lambda]} = \mathbb{E}_{\mathsf{\theta}_k}\!\big[ | \theta_{k} - \hat{\theta}_{m,k}^{[\lambda]} |^2 \big]$, $\psi_{m,\ell}^{t[\lambda]} = \mathbb{E}_{\mathsf{t}_{\ell}}\!\big[ | t_{\ell} - \hat{t}_{m,\ell}^{[\lambda]} |^2 \big]$ are the corresponding \acp{MSE}.
\newpage

In hand of the conditional \acp{PDF}, the extrinsic \ac{PDF} is obtained as \vspace{-1.5ex}
\begin{subequations}
\begin{align}
\prod_{i \neq m} p_{\tilde{\mathsf{z}}_{i,k}^{\theta[\lambda]} \mid \mathsf{\theta}_{k}}\left(\tilde{z}_{i,k}^{\theta[\lambda]} \mid \theta_{k}\right) &\propto \mathrm{exp}\bigg[ -\frac{|\theta_{k} - \bar{\theta}_{m,k}^{[\lambda]}|^2}{\bar{v}_{m,k}^{\theta[\lambda]}} \bigg], \\[-0.5ex]
\prod_{i \neq m} p_{\tilde{\mathsf{z}}_{i,\ell}^{t[\lambda]} \mid \mathrm{t}_{\ell}}\left(\tilde{z}_{i,\ell}^{t[\lambda]} \mid t_{\ell}\right) &\propto \mathrm{exp}\bigg[ -\frac{|t_{\ell} - \bar{t}_{m,\ell}^{[\lambda]}|^2}{\bar{v}_{m,\ell}^{t[\lambda]}} \bigg],
\end{align}
\label{eq:del_t_extr_PDF}
\vspace{-0.5ex}
\end{subequations}
where the corresponding extrinsic means and variances are defined as \vspace{-2ex}
\begin{subequations}
\label{eq:del_t_theta_extr_mean}
\begin{align}
\bar{\theta}_{m,k}^{[\lambda]} &= \bar{v}_{m,k}^{\theta[\lambda]} \bigg( \sum_{i \neq m} \frac{h^{\theta}_{i,k} \cdot \tilde{z}_{i,k}^{\theta[\lambda]}}{ \big(\sigma_{i,k}^{\theta[\lambda]}\big)^2} \bigg)\in \mathbb{R}, \label{eq:del_the_extr_mean}\\
\bar{t}_{m,\ell}^{[\lambda]} &= \bar{v}_{m,\ell}^{t[\lambda]} \bigg( \sum_{i \neq m} \frac{h^{t}_{i,\ell} \cdot \tilde{z}_{i,\ell}^{t[\lambda]}}{ \big(\sigma_{i,\ell}^{t[\lambda]}\big)^2} \bigg)\in \mathbb{R}, \label{eq:del_t_extr_mean}
\end{align}
\end{subequations}
\vspace{-2ex}
\begin{subequations}
\label{eq:del_t_theta_extr_var}
\begin{align}
\bar{v}_{m,k}^{\theta[\lambda]} &= \bigg( \sum_{i \neq m} \frac{|h^{\theta}_{i,k}|^2}{\big(\sigma_{i,k}^{\theta[\lambda]}\big)^2} \bigg)^{\!\!\!-1}  \!\!\!\!\in \mathbb{R}, \label{eq:del_the_extr_var} \\
\bar{v}_{m,\ell}^{t[\lambda]} &= \bigg( \sum_{i \neq m} \frac{|h^{t}_{i,\ell}|^2}{\big(\sigma_{m,\ell}^{t[\lambda]}\big)^2} \bigg)^{\!\!\!-1} \!\!\!\!\in \mathbb{R}. \label{eq:del_t_extr_var}
\end{align}
\end{subequations}

Then, the extrinsic means and variances are denoised with a Gaussian\footnote{If an alterative prior distribution of the position variables are assumed, \textit{i.e.,} uniform distribution, a different Bayes-optimal denoiser can be utilized.} prior to yield the new soft-replicas and \acp{MSE}, as \vspace{-2ex}
\begin{subequations}
\begin{equation}
\label{eq:del_t_soft_est}
\begin{aligned}
\check{\theta}_{m,k} &= \frac{\phi^{\theta} \cdot \bar{\theta}_{m,k}^{[\lambda]}}{\phi^{\theta} + \bar{v}_{m,k}^{\theta[\lambda]}} \in \mathbb{R}, 
& \check{t}_{m,\ell} &= \frac{\phi^{t} \cdot \bar{t}_{m,\ell}^{[\lambda]}}{\phi^{t} + \bar{v}_{m,\ell}^{t[\lambda]}} \in \mathbb{R}, \!\!\!\!\\
\end{aligned}
\end{equation}
\begin{equation}
\label{eq:del_t_est_mse}
\begin{aligned}
\check{\psi}_{m,k}^{\theta} &= \frac{\phi^{\theta} \cdot \bar{v}_{m,k}^{\theta[\lambda]}}{\phi^{\theta} + \bar{v}_{m,k}^{\theta[\lambda]}} \in \mathbb{R},
& \check{\psi}_{m,\ell}^{t} &= \frac{\phi^{t} \cdot \bar{v}_{m,\ell}^{t[\lambda]}}{\phi^{t} + \bar{v}_{m,\ell}^{t[\lambda]}} \in \mathbb{R}, \!\!\!\!\!\!\!\!\!\!
\end{aligned}
\end{equation}
\end{subequations}

\vspace{-0.5ex}
\noindent where $\phi^{\theta}$ and $\phi^{t}$ are the variance of the elements in $\boldsymbol{\theta}$ and $\boldsymbol{t}$.

\setcounter{footnote}{5}
\footnotetext{The effective channel powers are highly dependent on the sensor and anchor deployment structure, and for typical indoor sensing scenarios as illustrated in Figure \ref{fig:sys_mod_plot}, the anchor coordinates are of larger absolute value than the rigid body sensor coordinates, leading to the large power difference.}

Finally, the newly computed soft-replicas and \acp{MSE} are updated via damping to prevent early erroneous convergence to a local optima and error floor, which is described by
\begin{subequations}
\begin{equation}
\begin{aligned}
\hat{\theta}_{m, k}^{[\lambda+1]} &= \rho \hat{\theta}_{m, k}^{[\lambda]} + (1 - \rho) \check{\theta}_{m, k}^{[\lambda]},
\\
\hat{t}_{m, \ell}^{[\lambda+1]} &= \rho \hat{t}_{m, \ell}^{[\lambda]} + (1 - \rho) \check{t}_{m, \ell}^{[\lambda]}, \!\!\!\!\\
\end{aligned}
\end{equation}
\begin{equation}
\begin{aligned}
\psi_{m, k}^{\theta[\lambda+1]} &= \rho {\psi}_{m, k}^{\theta[\lambda]} + (1 - \rho) \check{\psi}_{m, k}^{\theta[\lambda]}, 
\\
\psi_{m, k}^{t[\lambda+1]} &= \rho {\psi}_{m, k}^{t[\lambda]} + (1 - \rho) \check{\psi}_{m, k}^{t[\lambda]}, 
\end{aligned}
\end{equation}
\label{eq:damped_update}
\end{subequations}
where $\rho \in [0,1]$ is the damping hyperparameter, and the superscript $(\cdot)^{[\lambda]}$ denotes the iterate at the $\lambda$-th \ac{GaBP} iteration.

After $\lambda_\mathrm{max}$ iterations of the \ac{GaBP} (or some convergence criteria), the final consensus estimates of the rotation and translation parameters are obtained as 
\begin{subequations}
\begin{align}
\tilde{\theta}_{k} &= \bigg( \sum_{m = 1}^{M} \frac{|h^{\theta}_{m,k}|^2}{\big(\sigma_{m,k}^{\theta[\lambda_\mathrm{max}]}\big)^2} \bigg)^{\!\!\!-1} \! \! \bigg( \sum_{m = 1}^{M} \frac{h^{\theta}_{m,k} \cdot \tilde{z}_{m,k}^{\theta[\lambda_\mathrm{max}]}}{ \big(\sigma_{m,k}^{\theta[\lambda_\mathrm{max}]}\big)^2} \bigg) \in \mathbb{R}, \nonumber \\[-2ex] \label{eq:del_t_final_est_theta} \\[-1ex]
\tilde{t}_{\ell} &= \bigg( \sum_{m = 1}^{M} \frac{|h^{t}_{m,\ell}|^2}{\big(\sigma_{m,\ell}^{t[\lambda_\mathrm{max}]}\big)^2} \bigg)^{\!\!\!-1} \!\bigg( \sum_{m = 1}^{M} \frac{h^{t}_{m,\ell} \cdot \tilde{z}_{m,\ell}^{t[\lambda_\mathrm{max}]}}{\big(\sigma_{m,\ell}^{t[\lambda_\mathrm{max}]}\big)^2} \,\bigg) \in \mathbb{R}. \nonumber \\[-2.5ex] \label{eq:del_t_final_est_t}
\end{align}
\label{eq:del_t_final_est}
\end{subequations}

\vspace{-3ex}
While the message passing rules elaborated are complete to yield the estimated rotation angles and translation vectors, due to the effective channel powers of $\boldsymbol{H}^{\theta}$ and $\boldsymbol{H}^{t}$ in equation \eqref{eq:delta_t_lin_syst} where the latter is typically much larger absolute positions of the anchors and sensors\footnotemark.
Such significant difference in effective channel powers lead to good estimation performance of the translation vector elements, but erroneous estimation performance of the rotation angles in a joint estimation described by the \ac{GaBP} procedure.

This behavior can also be intuitively understood by considering the illustration in Figure \ref{fig:RB_trans_plot}, where the rotation of the rigid body is expected to have a less prominent effect on the absolute sensor positions compared to the effect of the translation when the rotation angles are not too large, as assumed in the system formulation of Section \ref{sec:system_model}.

Then, in order to address the aforementioned error behavior of the rotation angle parameters $\boldsymbol{\theta}$, we propose an interference cancellation-based approach to remove the components corresponding to the translation of the sensors, and perform the \ac{GaBP} again only on the rotation angle parameters.
Namely, by using the estimated consensus translation vector $\tilde{\boldsymbol{t}} \triangleq [\tilde{t}_1, \tilde{t}_2, \tilde{t}_3]\trans \!\in \mathbb{R}^{3\times 1}$ obtained at the end of the \ac{GaBP} via equation \eqref{eq:del_t_final_est_t}, the interference-cancelled system is given by 
\begin{equation}
\label{eq:new_lin_syst}
\boldsymbol{z}_{n}' \triangleq \boldsymbol{z}_{n} - \boldsymbol{H}^{t} \tilde{\boldsymbol{t}} = \boldsymbol{H}^{\theta} \boldsymbol{\theta} + \boldsymbol{\xi}_{n} \in \mathbb{R}^{M \times 1}. \vspace{-1ex}
\end{equation}

The second \ac{GaBP} procedure to estimate the rotation angles $\boldsymbol{\theta}$ is identical to the message passing rules provided in equations \eqref{eq:del_t_soft_IC}-\eqref{eq:del_t_final_est}.

\vspace{-1ex}
{
\begin{algorithm}[H]
\caption{: Double \ac{GaBP} for \ac{RBL} Parameter Estimation}\label{alg:RBL_GaBP}
\hspace*{\algorithmicindent}
\begin{algorithmic}[1]
\vspace{-1.2ex}
\Statex \hspace{-4ex} \textbf{Input:} $\boldsymbol{z}_n \,(||\boldsymbol{s}_n||_2^2) \!~\forall n,  \boldsymbol{H}^{\theta}, \boldsymbol{H}^{t}, \phi^{\theta}, \phi^{t}, N_0, \lambda_\mathrm{max}, \rho$.
\Statex \hspace{-4ex}  \textbf{Output:} $\tilde{\theta}_{k}$ and $\tilde{t}_{\ell} ~\forall k,\ell$ (for all sensor nodes $\forall n$); \vspace{-1.5ex}
\Statex \hspace{-4.4ex} \hrulefill  \vspace{-0.3ex}
\Statex \hspace{-3.2ex} \textit{{Perform}} $\forall n, m, k, \ell:$ \vspace{0.25ex}
\State Initialise $\hat{\theta}_{m, k}^{[1]}$,  $\hat{t}_{m, \ell}^{[1]}$, $\psi_{m, k}^{\theta[1]}$, $\psi_{m, \ell}^{t[1]}$;
\For {$\lambda = 1$ to $\lambda_\mathrm{max}$}
\State \hspace{-3.5ex} Compute soft-\ac{IC} symbols $\tilde{z}_{m,k}^{\theta[\lambda]}$, $\tilde{z}_{m,\ell}^{t[\lambda]}$ via eq. \eqref{eq:del_t_soft_IC};
\State \hspace{-3.5ex} Compute cond. variances $\big(\!\sigma_{m,k}^{\theta[\lambda]}\big)^{\!2}\!\!\!, \big(\!\sigma_{m,\ell}^{t[\lambda]}\big)^{\!2}\!\!$ via eq. \eqref{eq:del_t_theta_var};
\State \hspace{-3.5ex} Compute extrinsic means $\bar{\theta}_{m,k}^{[\lambda]}$, $\bar{t}_{m,\ell}^{[\lambda]}$ via eq. \eqref{eq:del_t_theta_extr_mean};
\State \hspace{-3.5ex} Compute extrinsic variances $\bar{v}_{m,k}^{\theta[\lambda]}$, $\bar{v}_{m,\ell}^{t[\lambda]}$ via eq. \eqref{eq:del_t_theta_extr_var}; \vspace{-0.5ex}
\State \hspace{-3.5ex} Denoise the beliefs $\check{\theta}_{m,k}, \check{t}_{m,\ell}$ via eq. \eqref{eq:del_t_soft_est}; \vphantom{ $\check{x}_{m, k}^{[\lambda]}$}  \vspace{-0.5ex}
\State \hspace{-3.5ex} Denoise the error variances $\check{\psi}_{m,k}^{\theta}, \check{\psi}_{m,\ell}^{t}$ via eq. \eqref{eq:del_t_est_mse}; \vphantom{ $\check{x}_{m, k}^{[\lambda]}$}  \vspace{-0.5ex}
\State \hspace{-3.5ex} Update the soft-replicas with damping as in eq. \eqref{eq:damped_update}; \vphantom{ $\check{x}_{m, k}^{[\lambda]}$} 
\EndFor
\State Obtain final consensus estimates $\tilde{\theta}_{k}, \tilde{t}_{\ell}$ via eq. \eqref{eq:del_t_final_est};
\State Obtain interference-cancelled system via eq. \eqref{eq:new_lin_syst};
\For {$\lambda = 1$ to $\lambda_\mathrm{max}$}  \vspace{-0.2ex}
\State \hspace{-3.5ex} Compute soft-\ac{IC} symbols $\tilde{z}{'}_{\!\!m,k}^{\theta[\lambda]}$ via eq. \eqref{eq:new_lin_SIC};  \vspace{-0.5ex}
\State \hspace{-3.5ex} Compute conditional variances $\big(\sigma_{m,k}^{\theta[\lambda]}\big)^{2}$ via eq. \eqref{eq:new_lin_condvar};  \vspace{-0.5ex}
\State \hspace{-3.5ex} Compute extrinsic means $\bar{\theta}_{m,k}^{[\lambda]}$ via eq. \eqref{eq:del_the_extr_mean};  \vspace{-0.5ex}
\State \hspace{-3.5ex} Compute extrinsic variances $\bar{v}_{m,k}^{\theta[\lambda]}$ via eq. \eqref{eq:del_the_extr_var};  \vspace{-0.75ex}
\State \hspace{-3.5ex} Denoise the beliefs $\check{\theta}_{m,k}$ via eq. \eqref{eq:del_t_soft_est};  \vspace{-0.5ex} \vphantom{ $\check{x}_{m, k}^{[\lambda]}$}
\State \hspace{-3.5ex} Denoise the error variances $\check{\psi}_{m,k}^{\theta}$ via eq. \eqref{eq:del_t_est_mse}; \vphantom{ $\check{x}_{m, k}^{[\lambda]}$} \vspace{-0.75ex}
\State \hspace{-3.5ex} Update the soft-replicas with damping as in eq. \eqref{eq:damped_update}; \vphantom{ $\check{x}_{m, k}^{[\lambda]}$} 
\EndFor
\vspace{-0.25ex}
\State Obtain refined consensus estimates $\tilde{\theta}_{k}$ via eq. \eqref{eq:del_t_final_est_theta};
\vspace{-2ex}
\end{algorithmic} 
\hspace*{\algorithmicindent}
\end{algorithm}
}

After an \ac{IC} procedure to remove the effect of $\bm{t}$ from the system to change the message construction rules at the factor nodes for soft-\ac{IC} and conditional variance computations of equations \eqref{eq:del_t_soft_IC} and \eqref{eq:del_t_theta_var} into
\begin{equation}
\label{eq:new_lin_SIC}
\tilde{z}{'}_{\!\!m,k}^{\theta[\lambda]} = z'_{m} - \sum_{i \neq k} h^{\theta}_{m,i}\hat{\theta}_{m,i}^{[\lambda]} \in \mathbb{R}, \vspace{-1ex}
\end{equation}
\begin{equation}
\label{eq:new_lin_condvar}
\big(\sigma_{m,k}^{\theta[\lambda]}\big)^2 = \sum_{i \neq k} \left|h^{\theta}_{m,i}\right|^2\psi_{m,i}^{\theta[\lambda]} + N_{0} \in \mathbb{R}, \vspace{-1ex}
\end{equation}
while the remaining message passing rules on $\bm{\theta}$ by the variable nodes remain the same.

The resultant single-variable \ac{GaBP} on $\bm{\theta}$ is concatenated with the previous bivariate \ac{GaBP} to describe the complete estimation process of the rigid body transformation parameters $\boldsymbol{\theta}$ and $\boldsymbol{t}$, as summarized by Algorithm \ref{alg:RBL_GaBP}. \vspace{-1ex}


\vspace{2ex}
\section{Performance Analysis}
\label{sec:performance_analysis}

In this section, we present the simulation results to demonstrate effectiveness of the proposed bivariate \ac{GaBP} and interference cancellation-based refinement \ac{GaBP} concatenated in Algorithm \ref{alg:RBL_GaBP} for \ac{RBL}, directly in terms of the rotation and translation parameter estimation performance.
The performance is compared against the relevant \ac{SotA} \ac{RBL} solution, whose initial \ac{TOA}-based sensor position estimation is performed via the approach in \cite{MaICASSP2011} and the consequent \ac{RBL} parameter estimation proposed in \cite{ChenTSP2015}, which is performed via \ac{WLS}-based closed-form two-stage method, incorporating a \ac{DAC} approach.
Therefore, for a fair comparison, the initial sensor position information required for our proposed approach is also provided by the same \ac{SotA} \ac{TOA}-based method \cite{MaICASSP2011}, whose details has not been included in this article due to page limitations.\footnote{The extended derivations and simulation results, including a detailed complexity analysis, can be found in the journal version of this article \cite{Nic_6D_RBL}.}

The simulation setup is the scenario illustrated in Figure \ref{fig:sys_mod_plot}, where the rigid body is composed of $N = 8$ sensors positioned at the vertices of a unit cube at the origin, with sensor positions described by the conformation matrix given by
\begin{equation*}\label{eq:C_mat}
\boldsymbol{C} \!=\!\!
\resizebox{0.39 \textwidth}{!}{$\left[\begin{array}{lllllllc}
-0.5 &\! \phantom{-}0.5 &\! \phantom{-}0.5 &\! -0.5 &\! -0.5 &\! \phantom{-}0.5 &\! -0.5 &\! \phantom{-}0.5 \\
-0.5 &\!-0.5 &\! \phantom{-}0.5 &\!\phantom{-}0.5 &\! -0.5 &\! -0.5 & \!\phantom{-}0.5 & \!\phantom{-}0.5 \\
-0.5 &\! -0.5 &\! -0.5 &\! -0.5 & \!\phantom{-}0.5 &\! \phantom{-}0.5 &\! \phantom{-}0.5 &\! \phantom{-}0.5
\end{array}\right]
$} \!\!\in\! \mathbb{R}^{3 \times 8}\!,
\end{equation*}
and the $M = 8$ anchors are positioned at the vertices of a larger cube (\textit{i.e.,} an indoor scenario at the corners of the room), where the anchor conformation matrix $\boldsymbol{A} \in \mathbb{R}^{3\times 8}$ is given by
\begin{equation*}\label{eq:A_mat}
\boldsymbol{A} \!=\!\!
\resizebox{0.38 \textwidth}{!}{$
\left[\begin{array}{lllllllc}
-10 &\! \phantom{-}10 &\! \phantom{-}10 &\! -10 & \!-10 & \!\phantom{-}10 &\! -10 &\! \phantom{-}10 \\
-10 & \!-10 & \!\phantom{-}10 &\!\phantom{-}10 &\! -10 & \!-10 & \!\phantom{-}10 & \!\phantom{-}10 \\
-10 & \!-10 &\! -10 &\! -10 & \!\phantom{-}10 &\! \phantom{-}10 &\! \phantom{-}10 & \!\phantom{-}10
\end{array}\right]
$}  \!\!\in\! \mathbb{R}^{3 \times 8}\!.
\end{equation*}

The \ac{RBL} rotation angles $\theta_x, \theta_y, \theta_z$ follow a zero-mean Gaussian distribution of variance $\phi^{\theta} = 10$, and the \ac{RBL} translation vector elements also follow a zero-mean Gaussian distribution of variance $\phi^{t} = 5$.

The performance is assessed using the \ac{RMSE} defined as
\begin{equation}
\mathrm{RMSE} = \sqrt{\frac{1}{E}\textstyle\sum_{j=1}^{E}\|\hat{\boldsymbol{x}}^{[j]}-{\boldsymbol{x}}\|_2^2},
\end{equation}
where $\hat{\boldsymbol{x}}^{[j]}$ is the \ac{RBL} parameter vector (position, angle, or translation) estimated during the $j$-th Monte-Carlo simulation, $\boldsymbol{x}$ is the true \ac{RBL} parameter vector, and $E = 10^3$ is the total number of independent Monte-Carlo experiments used for the analysis, and is evaluated for different noise standard deviations $\sigma$ of equation \eqref{eq:range_model}, a.k.a. the range error in meters.

\begin{figure}[H]
\vspace{-1ex}
\centering
\includegraphics[width=0.97\columnwidth]{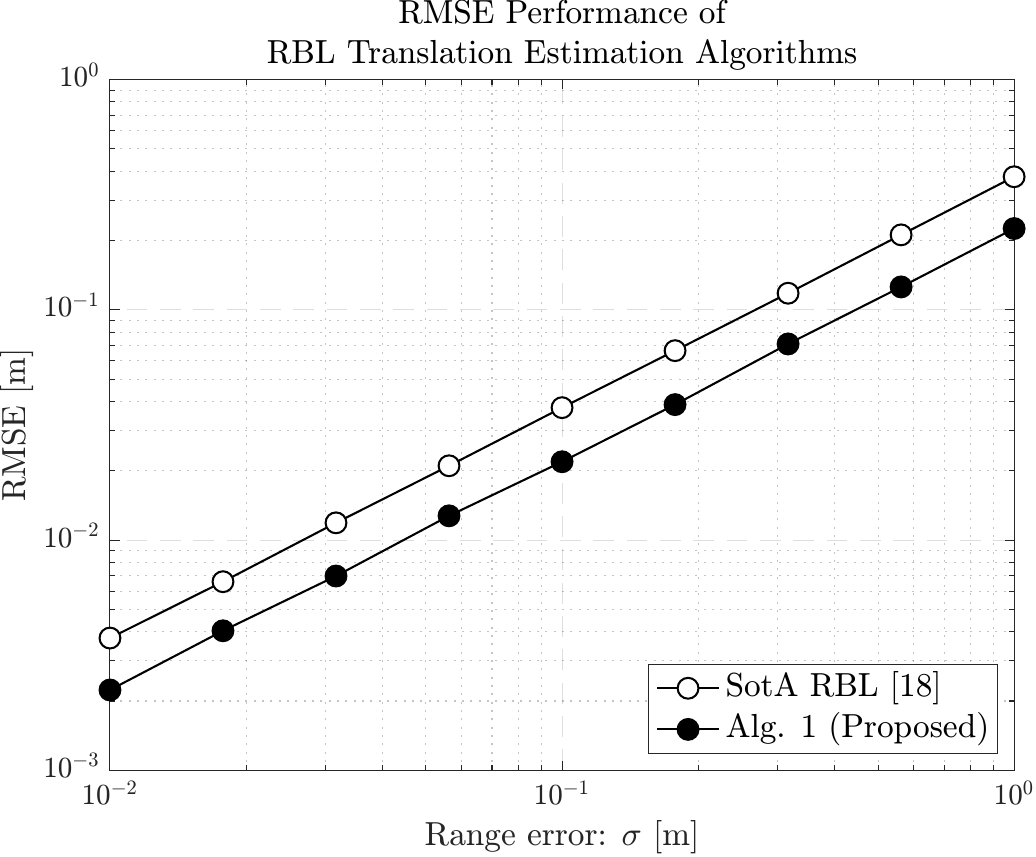}
\caption{\Acs{RMSE} of the proposed bivariate \ac{GaBP} algorithm for rigid body translation parameter estimation (Alg. \ref{alg:RBL_GaBP}) and the \ac{SotA} method of \cite{ChenTSP2015}, for various range errors $\sigma$.}
\label{fig:tran_rmse_plot}
\end{figure}

\begin{figure}[H]
\centering
\includegraphics[width=0.97\columnwidth]{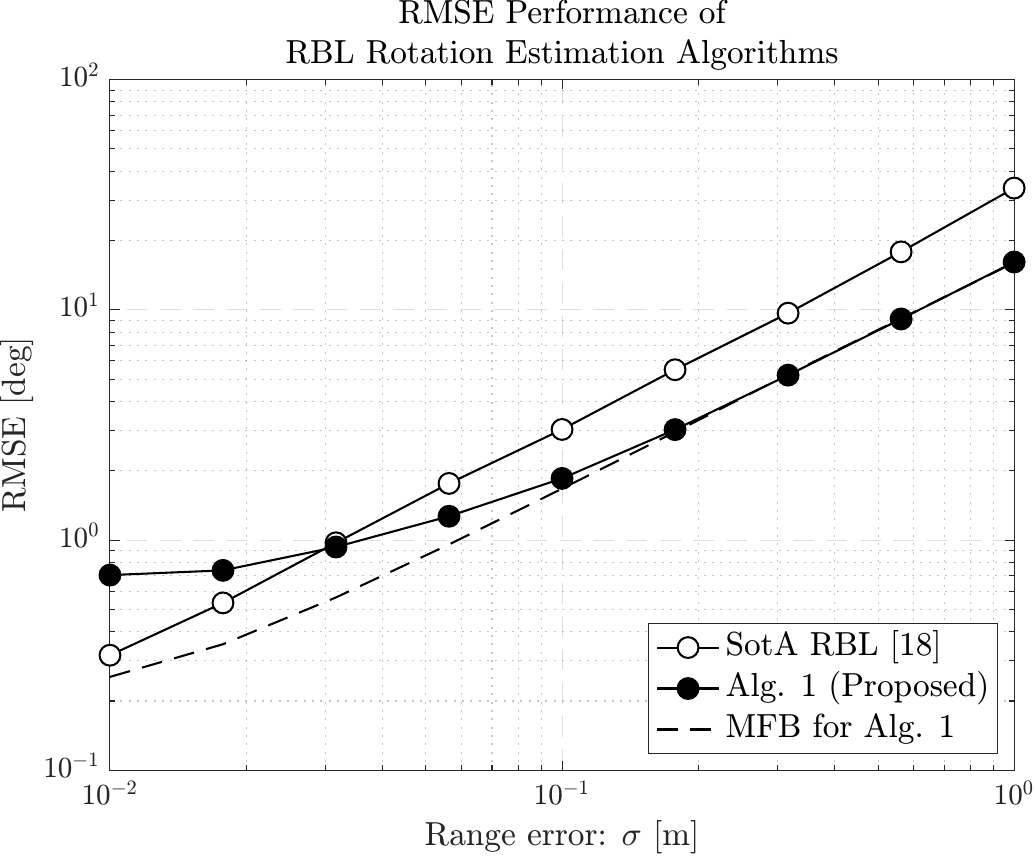}    
\vspace{-1ex}
\caption{\Ac{RMSE} of the proposed bivariate \ac{GaBP} algorithm for rigid body rotation angle parameter estimation (Alg. \ref{alg:RBL_GaBP}) and the \ac{SotA} method of \cite{ChenTSP2015}, for various range errors $\sigma$.}
\label{fig:rot_rmse_plot}
\end{figure}

The estimation performances of the rigid body translation parameters and rotation angle parameters have been illustrated in Figures \ref{fig:tran_rmse_plot} and \ref{fig:rot_rmse_plot}, respectively.

\vspace{-4.5ex}

As mentioned in Section \ref{sec:proposed}, the translation parameter can be effectively estimated via the proposed bivariate \ac{GaBP} of Algorithm \ref{alg:RBL_GaBP} in light of the channel power difference effect, which is highlighted by the result of Figure \ref{fig:tran_rmse_plot}.
It can be seen that the proposed bivariate \ac{GaBP} algorithm outperforms the \ac{SVD}-based \ac{SotA} approach \cite{ChenTSP2015} in terms of the rotation angle estimation, for all regimes of range error.

Finally, Figure \ref{fig:rot_rmse_plot} illustrates the estimation performance of the rotation angles of the rigid body transformation.
It can be observed that the estimation performance of the proposed concatenated \ac{GaBP} also exhibits superiority to the \ac{SotA} method, similarly to the behavior of the translation estimation performance.
However, due to the aforementioned channel power scaling effect which causes the noise power to be more prominent in the estimation of the variables, the \ac{GaBP} is shown to exhibit an error floor for small range error regimes.
Various methods exist to mitigate the error-floors \cite{LiTSP2024,takahashi2018design}, which are well-identified for message passing algorithms under non-ideal conditions, whose incorporation and extension is considered out of scope of this article.
Instead, the ideal behavior of the proposed algorithm is illustrated via the \ac{MFB} of the \ac{GaBP} algorithm, which shows the expected superiority over all noise ranges.

\vspace{-1ex}\section{Conclusion}
\label{sec:conclusion}
\vspace{-.95ex}
We presented a novel and efficient technique for solving the \ac{RBL} problem directly in terms of the \ac{3D} rotation and and translation parameters from known sensor positions, via a series of tailored \ac{GaBP} message passing estimators.
The \ac{RBL} system is first reformulated via the small-angle approximation to enable the construction of a bivariate \ac{GaBP} which is capable of directly estimating the rotation angles and the translation distances.
Then, a second interference cancellation-based second \ac{GaBP} to improve the performance of the rotation angle parameter estimation is incorporated, to mitigate the effect of imbalanced effective channel powers of the bivariate system.
The proposed concanated \ac{GaBP}-based \ac{RBL} method is shown to outperform the \ac{SotA} method in both the rotation and translation estimation performance, except for the appearance of an error-floor for the rotation angle estimation at low range error regimes.
Future works will aim to address the mitigation of such error-floor, enhanced robustness under diverse and non-ideal conditions of sensor deployment and conformations, via matrix completion and advanced belief propagation techniques.

\vspace{-0.78ex}

\end{document}